\documentclass[sigconf]{acmart}

 \usepackage{epsfig}
 \usepackage{algorithm} 
 \usepackage{epstopdf} 
\usepackage{algpseudocode}

\usepackage{tikz}
\usetikzlibrary{automata,positioning}
\usetikzlibrary{decorations.pathreplacing,decorations.markings,shapes.geometric}
\usetikzlibrary{shapes,arrows}
\usetikzlibrary{backgrounds,calc,positioning}

\usepackage{tabularx}       
\usepackage{float}          
\usepackage{booktabs}       

\usepackage{enumerate}

\def\be{ \begin{equation} }
\def\ee{ \end{equation} }
\def\bea{ \begin{eqnarray} }
\def\eea{ \end{eqnarray} }

\def\b0{{\bf 0}}

\catcode`,\active

\catcode`\,12

\theoremstyle{remark}

\renewcommand\footnotetextcopyrightpermission[1]{} 
\setcopyright{none}

\acmConference[SIGCOMM 2024 Sydney Australia]{}

\begin{document}

\title{Robust Zero Trust Architecture: Joint Blockchain based Federated learning and Anomaly Detection based Framework}

\author{Shiva Raj Pokhrel, Luxing Yang, Sutharshan Rajasegarar and Gang Li}
\thanks{School of IT, Deakin University, Australia, shiva.pokhrel@deakin.edu.au}

\renewcommand{\shortauthors}{Pokhrel et al.}

%
%

\begin{abstract}
    This paper introduces a robust zero-trust architecture (ZTA) tailored for the decentralized system that empowers efficient remote work and collaboration within IoT networks. Using blockchain-based federated learning principles, our proposed framework includes a robust aggregation mechanism designed to counteract malicious updates from compromised clients, enhancing the security of the global learning process. Moreover, secure and reliable trust computation is essential for remote work and collaboration. The robust ZTA framework integrates anomaly detection and trust computation, ensuring secure and reliable device collaboration in a decentralized fashion. We introduce an adaptive algorithm that dynamically adjusts to varying user contexts, using unsupervised clustering to detect novel anomalies, like zero-day attacks. To ensure a reliable and scalable trust computation, we develop an algorithm that dynamically adapts to varying user contexts by employing incremental anomaly detection and clustering techniques to identify and share local and global anomalies between nodes. Future directions include scalability improvements, Dirichlet process for advanced anomaly detection, privacy-preserving techniques, and the integration of post-quantum cryptographic methods to safeguard against emerging quantum threats.
\end{abstract}
\maketitle



\section{Introduction}\label{sec-intro}

In the post-quantum era, the landscape of remote computation, online collaboration, and work-from-home (WFH) has transformed dramatically. Leading cloud service providers, such as Amazon S3 and Azure Blobs, have made these capabilities standard, reflecting their importance. Astonishingly, 89\% of corporate data uploaded to the cloud is now associated with remote collaboration and file sharing. This shift requires a reevaluation of traditional network security concepts, especially perimeter-based security. The widespread adoption of cloud services and remote work has effectively dissolved physical network boundaries, calling for innovative security strategies to protect our increasingly virtualized work environments.

Traditional perimeter-based security approaches, such as enterprise firewalls, are increasingly inadequate. They struggle to: (i) effectively counteract malicious actors within the network perimeter, and (ii) provide sufficient protection for remote users, devices, and services outside the network perimeter. This shift underscores the urgent need for security architectures that are independent of both the perimeter and location.

The Zero Trust Access Architecture (ZTA) has emerged as a promising solution to this challenge \cite{rose2019zero}. ZTA fundamentally rethinks network security by assuming that the infrastructure is compromised and focuses on minimizing uncertainty in enforcing precise, least-privilege, per-request access decisions. It emphasizes situational awareness, dynamic authentication and authorization, fine-grained access control, small trust zones, and continuous trust evaluation. The literature on federated learning~\cite{pokhrel2020federated} is quite rich, including that to enable security; however, its utilization to operationalize ZTA is poorly studied in the literature. Pokhrel et al.~\cite{pokhrel2023towards} proposed a decentralized ZTA (dZTA); however, implementing these principles effectively and securely in large-scale enterprises or cyberphysical infrastructures, without compromising resource availability or dynamic control over network assets, remains a significant challenge that demands innovative and collaborative intelligence-driven solutions~\cite{pokhrel2024oZTA}. 
\begin{figure}[b]
    \centering
    \includegraphics[scale=0.24]{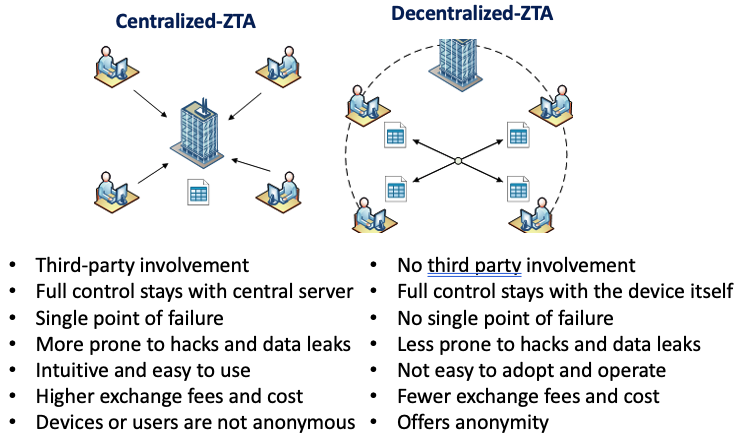}
    \caption{\rm Comparison of centralized and decentralized approaches for deploying ZTA, our approach considers distributed policy monitoring, trust computation, and continuous authentication to enhance security, scalability, and protection.}
    \label{fig:imp}
     \vspace{-6 mm}
\end{figure}

\subsection{Motivation and background} ZTA operates on the core principle of "\textbf{never trust, always verify}," applying this rigorously to both users and assets at a per request level of granularity. This approach presents a promising vision for network security, focusing on users, assets, and resources rather than solely on network access. In Figure~\ref{fig:imp}, we compare the pros and cons of centralized and decentralized approaches to ZTA deployment. Given the relentless attacks on central servers,\footnote{The Identity Theft Resource Center’s 2021 annual data breach report noted a record 1862 compromises, a 68\% increase from 2020. See \textit{www.idtheftcenter.org} for details.} ensuring security even when the server and certain devices are compromised is a critical concern. Several initiatives from academia and industry address this issue \cite{goh2003sirius, keybase, ardagna2015security, shraer2010venus, malarvizhi2014secure, pokhrel2024oZTA}.

Seminal solutions \cite{goh2003sirius, shraer2010venus, ardagna2015security} propose that remote devices encrypt and sign their files when sharing. However, powerful adversaries can:\\
i) Observe device identities and lists of WFH devices and learn their social relationship graphs.\\
ii) Launch active attacks, hide device updates (fork), or revert updates to earlier states (rollback), such as compromising medical profiles so patients are unaware of changes in their treatment procedures.
\subsection{Key ZT Challenges and Releated Work}
To address these issues, researchers have focused on two key ideas in the past decade: anonymity (masking user identities from the server) and verifiable consistency (allowing users to detect forks and rollbacks). However, current ZTA deployment strategies, including the NIST framework \cite{rose2019zero}, also rely heavily on centralized servers, presenting similar security challenges such as active attacks and single point of failure. Some approaches assume that adversaries are curious but not malicious \cite{maffei2015privacy, buccafurri2015accountability}, or they rely on a trusted party \cite{maffei2017maliciously} and split the server into components assuming one is honest \cite{7546541, khan2012anonymouscloud}. Yet, a persistent attacker can often compromise even these divided central servers, as shown in \cite{kim2015caelus, mazieres2002building}, because devices cannot detect fork attacks without trusting the server to some extent if they are not reliable online.

Therefore, in this paper, we identify and aim to solve a bottleneck challenge: \textit{ensuring robust security in a decentralized ZTA framework without relying on vulnerable central servers}.

A critical question that arises in this context is: Can strong security be achieved without depending on centralized trust in remote collaborative WFH systems? To address this, we designed a robust decentralized deployment framework for ZTA, focusing on distributed policy monitoring, trust computation, and continuous authentication. The primary objectives, as outlined in the NIST special publication on ZTA (August 2020), are to prevent data breaches and limit internal lateral movement. This paper aims to develop a fully decentralized and robust deployment of ZTA for trusted remote working and collaboration. Our framework is scalable, efficient, and provably secure, minimizing opportunities for lateral movement by both malicious and nonmalicious actors. This novel approach involves developing new algorithms for distributed trust computation and anomaly detection.

\section{Proposed Framework: Core Principles and Challenges}\label{sec:prinChall}
\begin{figure}[t]
    \centering
    \includegraphics[scale=0.3]{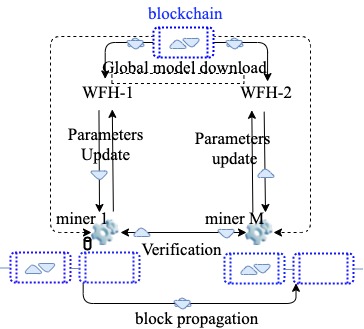}
    \caption{\rm Abstract View of the proposed Blockchain-enabled Federated Learning (BFL) Framework}
    \label{fig:imp2}
    \vspace{-5 mm}
\end{figure}

In recent years, federated learning has emerged as a compelling alternative to centralized approaches, offering significant advantages in terms of privacy and security \cite{pokhrel2020federated}. This paper introduces a novel blockchain-enabled federated learning (BFL) framework \cite{pokhrel2020federated, xu2022fair} to deploy a decentralized Zero Trust Architecture (dZTA). Our approach addresses the critical need for continuous monitoring and diagnosis inherent in ZTA implementations by developing a robust decentralized federated learning model, as shown in Figure~\ref{fig:imp2}, focusing on context awareness and anomaly detection for streaming data.

\subsection {Innovations and Benefits}
Our BFL framework uses blockchain technology to autonomously verify updates of the decentralized model, allowing real-time trust assessment and anomaly detection, which quickly addresses anomalies and prevents malicious interference. Innovative lifelong learning algorithms dynamically adapt to changing contexts, ensuring a reliable and secure trust computation for remote users. The integration of blockchain with federated learning ensures full decentralization, enhanced security, and robust model aggregation that maintains the integrity of the learning process despite potential corrupted updates. The details are discussed below.

\textit{Context Awareness and Anomaly Detection:} Our BFL framework leverages decentralized exchange of local model updates from users and devices. By utilizing blockchain technology, these updates are autonomously verified and used to assess trustworthiness and authorize access in real-time. This method ensures that any anomalies are detected and addressed quickly, preventing malicious clients from poisoning the learning process, and maintaining a self-motivated and reliable learning environment for all clients.

\textit{Decentralized Trust Computation:} Establishing trust in a remote working context, where users' physical locations, devices, network settings, and behaviors can dynamically change, requires innovative algorithms capable of lifelong learning, such as those of learning to harness~\cite{9444785}, studied in a different context. These algorithms must dynamically adapt to new contexts, ensuring reliable and secure trust computation even as conditions evolve.

\textit{Robust Federated Learning Framework:} The heart of our approach lies in the integration of blockchain with federated learning. By replacing the central server with decentralized blockchain technology, we ensure full decentralization and improved security. The use of smart contracts and blockchain transactions allows for autonomous enforcement of security policies, reducing the risk of single points of failure. Our design also incorporates robust model aggregation methods that can withstand corrupted updates from local devices, maintaining the integrity of the learning process.

It is worth noting that the proposed framework brings several benefits, including the following.

\begin{itemize}
    \item \textbf{Enhanced Security}: Autonomous verification and blockchain integration ensure decentralized trust and integrity.
    \item \textbf{Dynamic Trust Computation}: Lifelong learning algorithms provide reliable and adaptable trust scores.
    \item \textbf{Robust Model Aggregation}: Resilient aggregation methods maintain the integrity of the learning process despite possible corrupted updates.
\end{itemize}
\subsection{Underlying Analytic Understanding }
In the proposed BFL framework, the decentralized exchange of local model updates can be mathematically represented as follows. Each device $i$ in the network trains a local model $M_i$ on its data $D_i$ and generates an update $\Delta M_i$: $\Delta M_i = \text{Train}(M_i, D_i).$

These updates $\Delta M_i$ are then submitted to the blockchain. Let $B$ represent the blockchain, where each block $B_k$ contains a set of updates $\Delta M_{i_k}$ from various devices $i_k$:
\[
B_k = \{ \Delta M_{i_1}, \Delta M_{i_2}, \ldots, \Delta M_{i_n} \}.
\]
The updates are verified for trustworthiness using smart contracts $SC$. Each smart contract $SC_j$ performs verification on the update $\Delta M_i$ by checking consistency and integrity:
$\text{Verify}(\Delta M_i) = SC_j(\Delta M_i).$
 The verified updates are then aggregated to update the global model $M_G$:
\[
M_G = \text{Aggregate}(\{ \Delta M_{i_k} \mid \text{Verify}(\Delta M_{i_k}) = \text{True} \})
\]
Anomaly detection will be performed by analyzing the deviations of $\Delta M_i$ from the expected model update distribution. Let $E[\Delta M]$ be the expected update:
\[
\text{Anomaly}(\Delta M_i) = \| \Delta M_i - E[\Delta M] \| > \epsilon
\]
where $\epsilon$ is a predefined threshold.

To dynamically compute trust in a remote working context, the algorithm involves:\\
i. \textit{Trust Score Initialization}.
Each device $i$ starts with an initial trust score $T_i(0)$:
\[
T_i(0) = T_0.
\]
ii. \textit{Contextual Updates}.
Trust scores are updated based on the context $C_i(t)$ at time $t$, which includes physical location, device usage, network settings, and behavior:
\[
T_i(t+1) = f(T_i(t), C_i(t)).
\]
iii. \textit{Lifelong Learning}.
The trust computation adapts to new contexts using lifelong learning algorithms. Let $L$ be the lifelong learning algorithm:
\[
T_i(t+1) = L(T_i(t), C_i(t))
\]

The proposed robust federated learning framework can be viewed mathematically as follows.
Each device $i$ trains its local model $M_i$ and generates an update $\Delta M_i$ as before:
\[
\Delta M_i = \text{Train}(M_i, D_i).
\]
Updates $\Delta M_i$ are recorded on the blockchain, ensuring decentralization:
\[
B_k = \{ \Delta M_{i_1}, \Delta M_{i_2}, \ldots, \Delta M_{i_n} \}
\]
and smart contracts $SC_j$ enforce security policies by verifying updates:
\[
\text{Verify}(\Delta M_i) = SC_j(\Delta M_i),
\]
aggregation of model updates considers potential corrupted updates. Let $R$ be the robust aggregation function that is resilient to corrupted updates:
\[
M_G = R(\{ \Delta M_{i_k} \mid \text{Verify}(\Delta M_{i_k}) = \text{True} \}).
\]

Using these mathematical formulations, our proposed framework offers a scalable, efficient and secure solution for remote work and collaboration, advancing the field of cybersecurity.

\subsection {Addressing Impeding Challenges}
\textit{Decentralized Trust Computation and Anomaly Detection.}
The primary challenge in deploying a decentralized Zero Trust Architecture (dZTA) for secure remote work and collaboration is ensuring secure and reliable trust computation. This requires novel algorithms that can dynamically adapt to new contexts and detect anomalies using scalable measures and clustering techniques. The decentralized nature of this task, where trust evaluation and anomaly detection occur within local training clients, presents a significant research challenge.

\textit{Robust Federated Learning Framework.}
To achieve a robust dZTA, we must design a federated learning framework resilient to various types of update corruption, including malfunctions in low-cost hardware, corrupted data, and adversarial attacks. Our innovative model aggregation methods, implemented in a distributed ledger, ensure robustness even when multiple devices are compromised.

\textit{Technical and Strategic Benefits.}
Deploying a decentralized ZTA using our blockchain-enabled federated learning (BFL) framework offers profound benefits. It provides a cost-efficient, versatile solution for trusted remote collaboration, resilient to cyberattacks. The robust security architecture, mechanisms, protocols, and applications developed in this paper have significant implications for mission-critical operations such as battlefield surveillance and emergency response.

For Australia's digital security, this paper develops pathways to enhance the capability to secure sensitive and classified information while enabling flexible and remote working. The knowledge generated will create a competitive advantage for various sectors within the cybersecurity industry. In conclusion, our proposed dZTA framework, underpinned by blockchain-enabled federated learning, represents a significant advancement in the security of remote collaboration. We seek ways to address critical challenges and provide a scalable, efficient, and provably secure solution for modern network environments.

\section{Algorithms Design and Evaluation}\label{sec:AnaDes}

Based on the principles and challenges of ZTA mentioned earlier, we design the following two main tasks, viz. i) \emph{Anomaly Detection based Trust Computation} ii) \emph{Robust dZTA framework Design}  and associated deliverable followed by evaluation and future directions. 

\subsection{Anomaly Detection based Trust Computation}
\begin{algorithm}[!h]
\caption{Secure and Reliable Trust Computation}
\begin{algorithmic}[1]
\State \textbf{Initialize:} Set initial trust score $T_i(0) = T_0$ for each device $i$
\While{system is running}
    \For{each device $i$}
        \State \textbf{Collect context} $C_i(t)$ at $t$
        \State \textbf{Call Algorithm 2: Clustering-Based Anomaly \& Trust Computation}
        \State \textbf{Update trust score} $T_i(t+1)$: 
        \[
        T_i(t+1) = L(T_i(t), C_i(t))
        \]
        \State \textbf{Generate local model update} $\Delta M_i$: 
        \[
        \Delta M_i = \text{Train}(M_i, D_i)
        \]
        \State \textbf{Submit update} $\Delta M_i$ to blockchain: 
        \[
        B_k = B_k \cup \{\Delta M_i\}
        \]
        \State \textbf{Verify update} $\Delta M_i$ using smart contract $SC_j$: 
        \[
        \text{Verify}(\Delta M_i) = SC_j(\Delta M_i)
        \]
    \EndFor
    \State \textbf{Aggregate verified updates}:
    \[
    M_G = \text{Aggregate}(\{\Delta M_{i_k} \mid \text{Verify}(\Delta M_{i_k}) = \text{True} \})
    \]
    \State \textbf{Anomaly Detection}: For each update $\Delta M_i$, check:
    \[
    \text{Anomaly}(\Delta M_i) = \| \Delta M_i - E[\Delta M] \| > \epsilon
    \]
    \If{Anomaly detected}
        \State \textbf{Handle anomaly} (e.g., alert, isolate device)
    \EndIf
\EndWhile
\end{algorithmic}
\label{algoth1}
\vspace{-5 mm}
\end{algorithm}
Secure and reliable trust computation is fundamental for resilient remote working and collaboration. We develop an algorithm for trust computation in organizational security architecture that can dynamically adapt to new contexts where the user can vary in terms of physical location, device use, network location, behavior, etc. Based on established trust computation models, we construct new scalability measures and utilize unsupervised clustering algorithms to detect new types of anomalies (e.g. zero-day attacks) in streaming data. The details are shown in Algorithm~\ref{algoth1}.

See the steps in Algorithm~\ref{algoth1} for more details, where each device starts with an initial trust score \(T_i(0) = T_0\). The algorithm continuously collects contextual data \(C_i(t)\), such as obtaining contextual information such as physical location, device usage, network settings, and behavior at time $t$, for each device. Using a lifelong learning algorithm \(L\), it updates the trust score \(T_i(t+1)\) to adapt to new contexts. Each device trains its local model on its data and generates a model update \(\Delta M_i\). This update is submitted to the blockchain for decentralized verification. Smart contracts verify the integrity and trustworthiness of the updates submitted. Verified updates are aggregated to update the global model \(M_G\). Anomaly detection is performed by comparing updates with expected values and handling anomalies accordingly. By following these steps, the Algorithm~\ref{algoth1} ensures a secure and reliable trust computation in a decentralized Zero Trust Architecture, dynamically adapting to new contexts and effectively detecting anomalies.

\subsection{Clustering based Anomaly Detection}
To ensure reliable and scalable trust computation, we develop an algorithm that uses clustering-based scalability measures. The details are shown in Algorithm~\ref{alg:algo2}, which dynamically adapts to new contexts where user behavior, physical location, device usage, and network settings can vary. The algorithm employs incremental anomaly detection and utilizes clustering techniques, such as hyperspherical or hyperellipsoidal clusters \cite{rajasegarar2014hyperspherical,rajasegarar2014ellipsoidal}, to identify and share local and global anomalies between nodes. 

The Algorithm~\ref{alg:algo2} is explained below. Each device starts with an initial trust score \(T_i(0) = T_0\). The blockchain ledger \(B\) is initialized to record all relevant information, and a threshold \(\epsilon\) is defined for anomaly detection. A predefined interval is set for periodic updates. Each node clusters its flow data using a clustering algorithm. This process involves running the clustering algorithm on each node's data (flow records) to identify patterns and group similar data points. Each node identifies local anomalies based on distance metrics and computes local anomaly scores. This helps in detecting deviations from normal behavior at the individual node level.

Local cluster information is shared between nodes using the blockchain protocol. If the blockchain protocol is not feasible, point-to-point communication between neighboring nodes is used to ensure that local clustering information is distributed throughout the network. Global clusters are calculated from the local cluster information for each node, and global anomalies are identified from these global clusters. This step aggregates data from all nodes to provide a comprehensive view of the network's behavior.

\begin{algorithm}[H]
\caption{Clustering-Based Anomaly and Trust Computation}
\begin{algorithmic}[1]
\State \textbf{Initialization:} 
\State Set initial trust score $T_i(0) = T_0$ for each device $i$
\State Initialize blockchain ledger $B$
\State Define threshold $\epsilon$ for anomaly detection
\State Define a pre-defined interval for periodic updates

\While{ZTA as a Daemon Process}
    \For{each node $n$}
        \State \textbf{Data Clustering:}
        \State Cluster flow data at node $n$
        \State Run clustering on each node's data
                \State \textbf{Local Anomalies Detection:}
                \State Identify local anomalies
                \State Compute local anomaly scores
        \State \textbf{Information Sharing:}
        \State If Share local cluster information using blockchain
        \State Else use point-to-point communication
        \State \textbf{Global Clustering:}
        \State Compute global clusters from local cluster
        \State Identify global anomalies
        
        \State \textbf{Anomaly Score and Trust Value:}
        \State Define global anomaly scores
        \State Define trust values from [0, 1]
        
        \State \textbf{Periodic Updates:}
        \State Periodically repeat the above process 
        \State Update the node's trust values, store in blockchain
        
        \State \textbf{Trust Value Utilization:}
        \State Retrieves the stored trust values
    \EndFor
\EndWhile
\end{algorithmic}
\label{alg:algo2}
\end{algorithm}

Global anomaly scores are defined and trust values are computed based on these scores, with trust values ranging from [0, 1]. This quantifies the trustworthiness of each node based on the detected anomalies. The process is repeated periodically on a predefined interval or on demand basis. The trust values of the nodes are updated and stored on the blockchain to maintain an up-to-date record of the trust levels across the network. For trust values, ZTA retrieves the stored trust values to perform its operations. This ensures that the system can make informed decisions based on the latest trust assessments.

To achieve scalability, each node clusters its flow data, identifying local anomalies based on these clusters. The local cluster information is then shared between the nodes using the blockchain protocol. If blockchain implementation is not feasible, point-to-point communication is used instead. Global clusters are computed from this shared information, allowing the identification of global anomalies and the calculation of trust values for each node. This entire process is repeated at regular intervals or on demand, ensuring that trust values are continually updated and stored in the blockchain. This maintains an up-to-date record of trust levels across the network. When the ZTA requires trust values, it retrieves the stored trust values from the blockchain. This ensures that the system can make informed and secure decisions based on the latest trust assessments.

To ensure the highest level of security, the accuracy of anomaly detection can be compared with traditional centralized schemes. This rigorous comparison should demonstrate the effectiveness of our decentralized approach in identifying and mitigating potential threats. Additionally, by sharing only cluster information instead of raw node data, we can significantly reduce communication overhead. This approach not only improves efficiency, but also ensures that sensitive data remain secure and private.


{

\subsection{Evaluation and Future Direction}

We used the implementation of BFL~\cite{xu2022fair} and evaluated the performance of the proposed framework using the well-known MNIST classification dataset. For tractability, we adopted the same parameter setting as that of~\cite{xu2022fair} and, in our experiments, there are \(100\) devices and 2 miners that cooperate to perform the trust computation. With 5 epochs of the local model and 10 in a batch at a learning rate of 0.01, we have the following observations and findings.
\begin{figure}
    \centering
    \includegraphics[scale=.3]{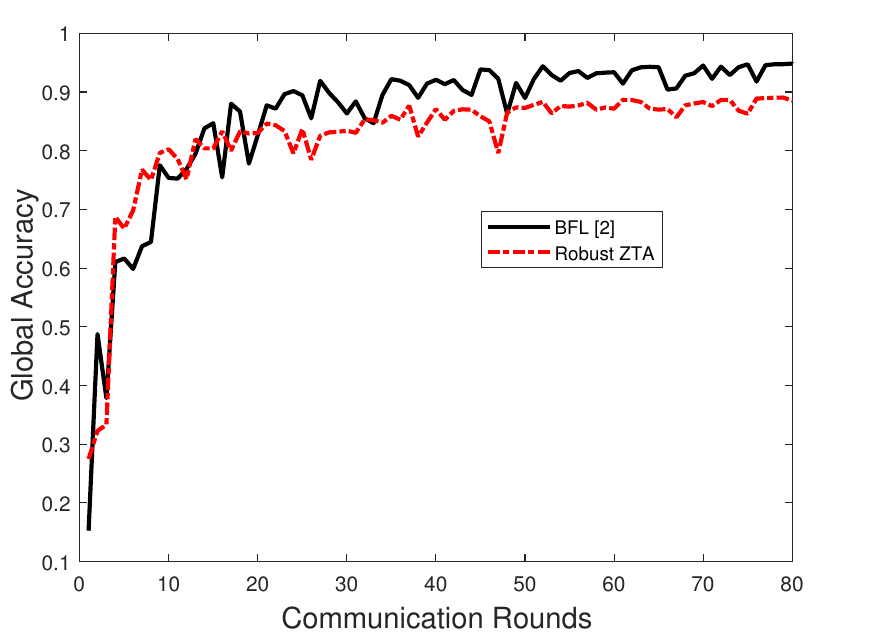}
    \caption{\rm Comparison of the temporal evolution of accuracy of the global model of the proposed Robust ZTA over communication rounds with that of BFL~\cite{pokhrel2020federated}.}
    \label{fig:l}
    \vspace{-5 mm}
\end{figure}

\begin{figure}
    \centering
    \includegraphics[scale=.3]{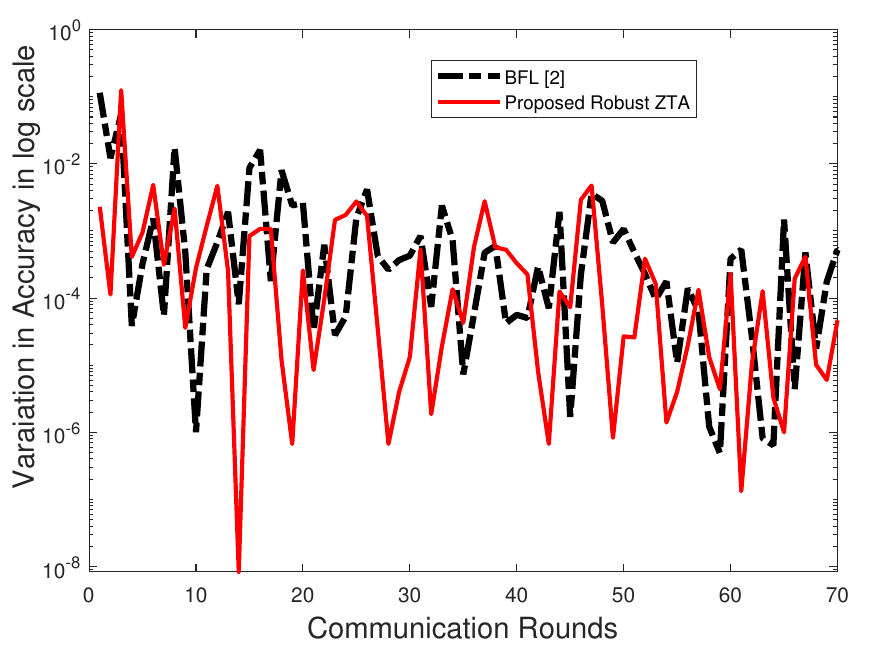}
    \caption{\rm Variation of the accuracy of the proposed Robust ZTA over communication rounds with that of BFL~\cite{pokhrel2020federated}.}
    \label{fig:2l}
     \vspace{-5 mm}
\end{figure}

\begin{figure}
    \centering
    \includegraphics[scale=.3]{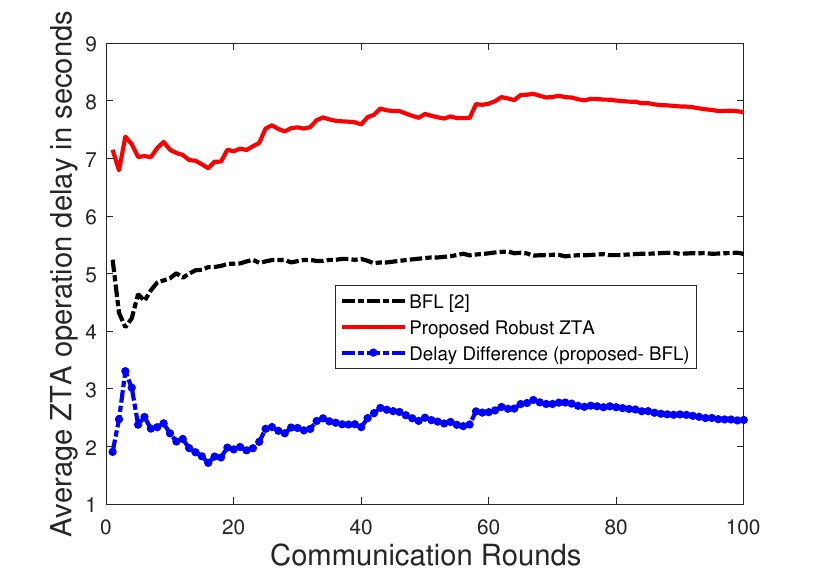}
    \caption{\rm System level delay comparison of the proposed Robust ZTA over communication rounds with that of BFL~\cite{pokhrel2020federated}.}
    \label{fig:3l}
     \vspace{-6 mm}
\end{figure}

We first consider scenarios where devices might fail, such as during a power outage, with and without blockchain support. Without blockchain, the failure of a central server significantly impacts the accuracy of learning. Our proposed Robust Zero Trust Architecture (ZTA) based on the BFL algorithm, which incorporates blockchain support, outperforms the baseline FL algorithms that lack blockchain support. In the Google FL approach, an unexpected central server failure severely degrades the accuracy of the learning. In contrast, our framework leverages blockchain to enable devices to act as global aggregators, maintaining global aggregation even if some devices fail.

In FL, the complexity and diversity of network architectures can lead to malicious updates from participating devices. The decoupling of the local training process from the server opens the possibility of poisoning attacks. In our second experiment, we assume the presence of malicious devices within the network attempting to degrade the performance of the local model and contaminate the global model. We simulated 10 malicious devices and evaluated the performance of our proposed aggregation mechanism. The results, shown in Figures~\ref{fig:l},\ref{fig:2l}, and \ref{fig:3l}, demonstrate that the performance of the proposed robust ZTA is comparable to that of BFL in terms of the evolution of the precision of the global model. This is a significant achievement, although it requires extensive validation in real-world scenarios. By examining the variation in accuracy and gauging the oscillation in Figure~\ref{fig:2l} of the robust ZTA proposed, we found it to be slightly more than BFL but still within a tolerable range. However, we discover a significant trade-off of around "2 seconds" in Figure~\ref{fig:3l} in terms of system-level delay comparison of the proposed Robust ZTA over communication rounds with that of BFL, which requires major evaluation in the future by implementing the proposed Dirichlet process for incremental anomaly detection.

\begin{figure}
    \centering
    \includegraphics[scale=0.65]{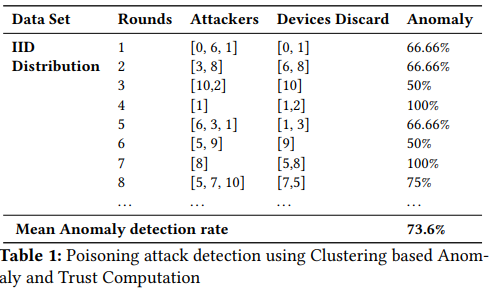}
    \label{fig:enter-label}
     \vspace{-7 mm}
\end{figure}


We emulate a discard strategy for attack devices to combat malicious activity, specifically targeting 10 malicious nodes that modify local gradients to skew the global model. With 10 indexed devices, in each communication round, we randomly designate 1 to 3 clients as malicious nodes, executing a total of 20 rounds, as detailed in Table 1. We can see that even with the vast majority of nodes remaining honest, the impact of poisoning attacks is evident. The modified local gradients stand out as anomalies compared to the normal ones. However, as the number of malicious nodes increases, some forged gradients manage to bypass the detection mechanism, leading to a decrease in the detection rate. This occurs because obvious anomalies can overshadow subtler ones, making them harder to detect.

Future research could explore the following directions to further enhance the proposed framework:\\
 \textbf{Scalability Improvements:} Investigate advanced clustering and aggregation techniques to enhance the scalability of the framework, particularly in large-scale deployments with thousands of devices.\\
 \textbf{Advanced Anomaly Detection:} Develop more sophisticated anomaly detection algorithms that can identify subtle and complex attack patterns, leveraging machine learning and deep learning techniques.\\
 \textbf{Privacy-Preserving Techniques:} Integrate privacy preserving mechanisms, such as differential privacy, to further protect sensitive data while maintaining high model accuracy and security.\\
 \textbf{Cross-Domain Applications:} Extend the application of the framework to other domains beyond classification tasks, such as regression, reinforcement learning, and natural language processing.\\
 \textbf{Real-World Deployments:} Conduct real-world deployments and case studies to evaluate the practical applicability and effectiveness of the framework in various industries and environments.

As quantum computing continues to advance, ensuring post-quantum security within our dZTA framework is crucial. Post-quantum research directions should include:\\
\textbf{Post-Quantum Cryptography:} Integrate post-quantum cryptographic algorithms to safeguard blockchain transactions and communications against quantum attacks. This will involve exploring quantum-resistant algorithms, such as lattice-based, hash-based, and multivariate polynomial cryptography.\\
 \textbf{Quantum-Resilient Consensus Mechanisms:} Develop and implement consensus mechanisms that are resistant to quantum computing threats, ensuring the integrity and security of the decentralized ledger.\\
 \textbf{Quantum Machine Learning:} Investigate the potential of quantum machine learning to enhance the efficiency and effectiveness of anomaly detection and trust computation in the federated learning~\cite{pokhrel2024quantum,hdaib2024quantum} setup for ZTA.\\
 \textbf{Hybrid Quantum-Classical Solutions:} Explore hybrid solutions that leverage both classical and quantum computing resources to optimize performance and security in real-world deployments of the dZTA framework.\\
 \textbf{Quantum Network Security:} Develop protocols for secure quantum communications and quantum key distribution (QKD) to enhance the overall security of the network against future quantum threats.

By addressing these future and post-quantum directions, we can further solidify the robustness and applicability of the dZTA framework, making it a versatile and essential solution for secure and reliable remote collaboration in diverse and evolving technological landscapes.
\section{Concluding Remarks }


In this paper, we have presented a robust decentralized Zero Trust Architecture  framework underpinned by a blockchain-enabled federated learning (BFL) approach. Our framework addresses critical challenges in secure remote work and collaboration by ensuring reliable trust computation, dynamic context adaptation, and robust anomaly detection. By leveraging blockchain technology, our proposed method mitigates the risks associated with central server failures and malicious model updates, providing a resilient and secure learning environment. Experimental results demonstrate the superior performance of our BFL algorithm compared to traditional federated learning approaches, particularly in scenarios involving device failures and poisoning attacks.


\bibliographystyle{ACM-Reference-Format}
\bibliography{yourbib}


\begin{thebibliography}{22}


\ifx \showCODEN    \undefined \def \showCODEN     #1{\unskip}     \fi
\ifx \showDOI      \undefined \def \showDOI       #1{#1}\fi
\ifx \showISBNx    \undefined \def \showISBNx     #1{\unskip}     \fi
\ifx \showISBNxiii \undefined \def \showISBNxiii  #1{\unskip}     \fi
\ifx \showISSN     \undefined \def \showISSN      #1{\unskip}     \fi
\ifx \showLCCN     \undefined \def \showLCCN      #1{\unskip}     \fi
\ifx \shownote     \undefined \def \shownote      #1{#1}          \fi
\ifx \showarticletitle \undefined \def \showarticletitle #1{#1}   \fi
\ifx \showURL      \undefined \def \showURL       {\relax}        \fi
\providecommand\bibfield[2]{#2}
\providecommand\bibinfo[2]{#2}
\providecommand\natexlab[1]{#1}
\providecommand\showeprint[2][]{arXiv:#2}

\bibitem[\protect\citeauthoryear{??}{key}{[n.d.]}]%
        {keybase}
 \bibinfo{year}{[n.d.]}\natexlab{}.
\newblock \showarticletitle{Keybase.io. https://keybase:io/}.
\newblock  (\bibinfo{year}{[n.\,d.]}).
\newblock


\bibitem[\protect\citeauthoryear{Ardagna et~al\mbox{.}}{Ardagna
  et~al\mbox{.}}{2015}]%
        {ardagna2015security}
\bibfield{author}{\bibinfo{person}{Claudio~A Ardagna} {et~al\mbox{.}}}
  \bibinfo{year}{2015}\natexlab{}.
\newblock \showarticletitle{From security to assurance in the cloud: A survey}.
\newblock \bibinfo{journal}{\emph{ACM Computing Surveys (CSUR)}}
  \bibinfo{volume}{48}, \bibinfo{number}{1} (\bibinfo{year}{2015}),
  \bibinfo{pages}{1--50}.
\newblock


\bibitem[\protect\citeauthoryear{Buccafurri, Lax, Nicolazzo, and
  Nocera}{Buccafurri et~al\mbox{.}}{2015}]%
        {buccafurri2015accountability}
\bibfield{author}{\bibinfo{person}{Francesco Buccafurri},
  \bibinfo{person}{Gianluca Lax}, \bibinfo{person}{Serena Nicolazzo}, {and}
  \bibinfo{person}{Antonino Nocera}.} \bibinfo{year}{2015}\natexlab{}.
\newblock \showarticletitle{Accountability-preserving anonymous delivery of
  cloud services}. In \bibinfo{booktitle}{\emph{International Conference on
  Trust and Privacy in Digital Business}}. Springer, \bibinfo{pages}{124--135}.
\newblock


\bibitem[\protect\citeauthoryear{Goh, Shacham, Modadugu, and Boneh}{Goh
  et~al\mbox{.}}{[n.d.]}]%
        {goh2003sirius}
\bibfield{author}{\bibinfo{person}{Eu-Jin Goh}, \bibinfo{person}{Hovav
  Shacham}, \bibinfo{person}{Nagendra Modadugu}, {and} \bibinfo{person}{Dan
  Boneh}.} \bibinfo{year}{[n.d.]}\natexlab{}.
\newblock \showarticletitle{SiRiUS: Securing Remote Untrusted Storage.}
  Citeseer.
\newblock


\bibitem[\protect\citeauthoryear{Hdaib, Rajasegarar, and Pan}{Hdaib
  et~al\mbox{.}}{2024}]%
        {hdaib2024quantum}
\bibfield{author}{\bibinfo{person}{Moe Hdaib}, \bibinfo{person}{Sutharshan
  Rajasegarar}, {and} \bibinfo{person}{Lei Pan}.}
  \bibinfo{year}{2024}\natexlab{}.
\newblock \showarticletitle{Quantum deep learning-based anomaly detection for
  enhanced network security}.
\newblock \bibinfo{journal}{\emph{Quantum Machine Intelligence}}
  \bibinfo{volume}{6}, \bibinfo{number}{1} (\bibinfo{year}{2024}),
  \bibinfo{pages}{26}.
\newblock


\bibitem[\protect\citeauthoryear{Karapanos, Filios, Popa, and Capkun}{Karapanos
  et~al\mbox{.}}{2016}]%
        {7546541}
\bibfield{author}{\bibinfo{person}{Nikolaos Karapanos},
  \bibinfo{person}{Alexandros Filios}, \bibinfo{person}{Raluca~Ada Popa}, {and}
  \bibinfo{person}{Srdjan Capkun}.} \bibinfo{year}{2016}\natexlab{}.
\newblock \showarticletitle{Verena: End-to-End Integrity Protection for Web
  Applications}. In \bibinfo{booktitle}{\emph{2016 IEEE Symposium on Security
  and Privacy (SP)}}. \bibinfo{pages}{895--913}.
\newblock
\urldef\tempurl%
\url{https://doi.org/10.1109/SP.2016.58}
\showDOI{\tempurl}


\bibitem[\protect\citeauthoryear{Khan and Hamlen}{Khan and Hamlen}{2012}]%
        {khan2012anonymouscloud}
\bibfield{author}{\bibinfo{person}{Safwan~Mahmud Khan} {and}
  \bibinfo{person}{Kevin~W Hamlen}.} \bibinfo{year}{2012}\natexlab{}.
\newblock \showarticletitle{AnonymousCloud: A data ownership privacy provider
  framework in cloud computing}. In \bibinfo{booktitle}{\emph{2012 IEEE 11th
  International Conference on Trust, Security and Privacy in Computing and
  Communications}}. IEEE, \bibinfo{pages}{170--176}.
\newblock


\bibitem[\protect\citeauthoryear{Kim and Lie}{Kim and Lie}{2015}]%
        {kim2015caelus}
\bibfield{author}{\bibinfo{person}{Beom~Heyn Kim} {and} \bibinfo{person}{David
  Lie}.} \bibinfo{year}{2015}\natexlab{}.
\newblock \showarticletitle{Caelus: Verifying the consistency of cloud services
  with battery-powered devices}. In \bibinfo{booktitle}{\emph{2015 IEEE
  Symposium on Security and Privacy}}. IEEE, \bibinfo{pages}{880--896}.
\newblock


\bibitem[\protect\citeauthoryear{Maffei, Malavolta, Reinert, and
  Schr{\"o}der}{Maffei et~al\mbox{.}}{2015}]%
        {maffei2015privacy}
\bibfield{author}{\bibinfo{person}{Matteo Maffei}, \bibinfo{person}{Giulio
  Malavolta}, \bibinfo{person}{Manuel Reinert}, {and}
  \bibinfo{person}{Dominique Schr{\"o}der}.} \bibinfo{year}{2015}\natexlab{}.
\newblock \showarticletitle{Privacy and access control for outsourced personal
  records}. In \bibinfo{booktitle}{\emph{2015 IEEE Symposium on Security and
  Privacy}}. IEEE, \bibinfo{pages}{341--358}.
\newblock


\bibitem[\protect\citeauthoryear{Maffei, Malavolta, Reinert, and
  Schr{\"o}der}{Maffei et~al\mbox{.}}{2017}]%
        {maffei2017maliciously}
\bibfield{author}{\bibinfo{person}{Matteo Maffei}, \bibinfo{person}{Giulio
  Malavolta}, \bibinfo{person}{Manuel Reinert}, {and}
  \bibinfo{person}{Dominique Schr{\"o}der}.} \bibinfo{year}{2017}\natexlab{}.
\newblock \showarticletitle{Maliciously secure multi-client ORAM}. In
  \bibinfo{booktitle}{\emph{International Conference on Applied Cryptography
  and Network Security}}. Springer, \bibinfo{pages}{645--664}.
\newblock


\bibitem[\protect\citeauthoryear{Malarvizhi, Sujana, and Revathi}{Malarvizhi
  et~al\mbox{.}}{2014}]%
        {malarvizhi2014secure}
\bibfield{author}{\bibinfo{person}{M Malarvizhi},
  \bibinfo{person}{J~Angela~Jennifa Sujana}, {and} \bibinfo{person}{T
  Revathi}.} \bibinfo{year}{2014}\natexlab{}.
\newblock \showarticletitle{Secure file sharing using cryptographic techniques
  in cloud}. In \bibinfo{booktitle}{\emph{2014 International Conference on
  Green Computing Communication and Electrical Engineering (ICGCCEE)}}. IEEE,
  \bibinfo{pages}{1--6}.
\newblock


\bibitem[\protect\citeauthoryear{Mazieres and Shasha}{Mazieres and
  Shasha}{2002}]%
        {mazieres2002building}
\bibfield{author}{\bibinfo{person}{David Mazieres} {and}
  \bibinfo{person}{Dennis Shasha}.} \bibinfo{year}{2002}\natexlab{}.
\newblock \showarticletitle{Building secure file systems out of Byzantine
  storage}. In \bibinfo{booktitle}{\emph{Proceedings of the twenty-first annual
  symposium on Principles of distributed computing}}.
  \bibinfo{pages}{108--117}.
\newblock


\bibitem[\protect\citeauthoryear{Pokhrel}{Pokhrel}{2024}]%
        {pokhrel2024oZTA}
\bibfield{author}{\bibinfo{person}{Shiva~Raj Pokhrel}.}
  \bibinfo{year}{2024}\natexlab{}.
\newblock \showarticletitle{Orbital ZTA: Secure Satellite Communication
  Networks with Zero Trust Architecture}.
\newblock \bibinfo{journal}{\emph{ACM SIGCOMM Computer Communication Review}}
  \bibinfo{volume}{48}, \bibinfo{number}{1}, \bibinfo{pages}{0--4}.
\newblock


\bibitem[\protect\citeauthoryear{Pokhrel and Choi}{Pokhrel and Choi}{2020}]%
        {pokhrel2020federated}
\bibfield{author}{\bibinfo{person}{Shiva~Raj Pokhrel} {and}
  \bibinfo{person}{Jinho Choi}.} \bibinfo{year}{2020}\natexlab{}.
\newblock \showarticletitle{Federated learning with blockchain for autonomous
  vehicles: Analysis and design challenges}.
\newblock \bibinfo{journal}{\emph{IEEE Transactions on Communications}}
  \bibinfo{volume}{68}, \bibinfo{number}{8} (\bibinfo{year}{2020}),
  \bibinfo{pages}{4734--4746}.
\newblock


\bibitem[\protect\citeauthoryear{Pokhrel, Li, Doss, and Nepal}{Pokhrel
  et~al\mbox{.}}{2023}]%
        {pokhrel2023towards}
\bibfield{author}{\bibinfo{person}{Shiva~Raj Pokhrel}, \bibinfo{person}{Gang
  Li}, \bibinfo{person}{Robin Doss}, {and} \bibinfo{person}{Surya Nepal}.}
  \bibinfo{year}{2023}\natexlab{}.
\newblock \showarticletitle{Towards Decentralized Operationalization of Zero
  Trust Architecture}.
\newblock \bibinfo{journal}{\emph{Authorea Preprints}} (\bibinfo{year}{2023}).
\newblock


\bibitem[\protect\citeauthoryear{Pokhrel and Walid}{Pokhrel and Walid}{2023}]%
        {9444785}
\bibfield{author}{\bibinfo{person}{Shiva~Raj Pokhrel} {and}
  \bibinfo{person}{Anwar Walid}.} \bibinfo{year}{2023}\natexlab{}.
\newblock \showarticletitle{Learning to Harness Bandwidth With Multipath
  Congestion Control and Scheduling}.
\newblock \bibinfo{journal}{\emph{IEEE Transactions on Mobile Computing}}
  \bibinfo{volume}{22}, \bibinfo{number}{2} (\bibinfo{year}{2023}),
  \bibinfo{pages}{996--1009}.
\newblock
\urldef\tempurl%
\url{https://doi.org/10.1109/TMC.2021.3085598}
\showDOI{\tempurl}


\bibitem[\protect\citeauthoryear{Pokhrel, Yash, Kua, Li, and Pan}{Pokhrel
  et~al\mbox{.}}{2024}]%
        {pokhrel2024quantum}
\bibfield{author}{\bibinfo{person}{Shiva~Raj Pokhrel}, \bibinfo{person}{Naman
  Yash}, \bibinfo{person}{Jonathan Kua}, \bibinfo{person}{Gang Li}, {and}
  \bibinfo{person}{Lei Pan}.} \bibinfo{year}{2024}\natexlab{}.
\newblock \showarticletitle{Quantum Federated Learning Experiments in the Cloud
  with Data Encoding}.
\newblock \bibinfo{journal}{\emph{arXiv preprint arXiv:2405.00909}}
  (\bibinfo{year}{2024}).
\newblock


\bibitem[\protect\citeauthoryear{Rajasegarar, Gluhak, Imran, Nati, Moshtaghi,
  Leckie, and Palaniswami}{Rajasegarar et~al\mbox{.}}{2014a}]%
        {rajasegarar2014ellipsoidal}
\bibfield{author}{\bibinfo{person}{Sutharshan Rajasegarar},
  \bibinfo{person}{Alexander Gluhak}, \bibinfo{person}{Muhammad~Ali Imran},
  \bibinfo{person}{Michele Nati}, \bibinfo{person}{Masud Moshtaghi},
  \bibinfo{person}{Christopher Leckie}, {and} \bibinfo{person}{Marimuthu
  Palaniswami}.} \bibinfo{year}{2014}\natexlab{a}.
\newblock \showarticletitle{Ellipsoidal neighbourhood outlier factor for
  distributed anomaly detection in resource constrained networks}.
\newblock \bibinfo{journal}{\emph{Pattern recognition}} \bibinfo{volume}{47},
  \bibinfo{number}{9} (\bibinfo{year}{2014}), \bibinfo{pages}{2867--2879}.
\newblock


\bibitem[\protect\citeauthoryear{Rajasegarar, Leckie, and
  Palaniswami}{Rajasegarar et~al\mbox{.}}{2014b}]%
        {rajasegarar2014hyperspherical}
\bibfield{author}{\bibinfo{person}{Sutharshan Rajasegarar},
  \bibinfo{person}{Christopher Leckie}, {and} \bibinfo{person}{Marimuthu
  Palaniswami}.} \bibinfo{year}{2014}\natexlab{b}.
\newblock \showarticletitle{Hyperspherical cluster based distributed anomaly
  detection in wireless sensor networks}.
\newblock \bibinfo{journal}{\emph{J. Parallel and Distrib. Comput.}}
  \bibinfo{volume}{74}, \bibinfo{number}{1} (\bibinfo{year}{2014}),
  \bibinfo{pages}{1833--1847}.
\newblock


\bibitem[\protect\citeauthoryear{Rose, Borchert, Mitchell, and Connelly}{Rose
  et~al\mbox{.}}{2020}]%
        {rose2019zero}
\bibfield{author}{\bibinfo{person}{Scott Rose}, \bibinfo{person}{Oliver
  Borchert}, \bibinfo{person}{Stu Mitchell}, {and} \bibinfo{person}{Sean
  Connelly}.} \bibinfo{year}{2020}\natexlab{}.
\newblock \bibinfo{booktitle}{\emph{Zero trust architecture}}.
\newblock \bibinfo{type}{{T}echnical {R}eport}. \bibinfo{institution}{National
  Institute of Standards and Technology}.
\newblock


\bibitem[\protect\citeauthoryear{Shraer, Cachin, Cidon, Keidar, Michalevsky,
  and Shaket}{Shraer et~al\mbox{.}}{2010}]%
        {shraer2010venus}
\bibfield{author}{\bibinfo{person}{Alexander Shraer},
  \bibinfo{person}{Christian Cachin}, \bibinfo{person}{Asaf Cidon},
  \bibinfo{person}{Idit Keidar}, \bibinfo{person}{Yan Michalevsky}, {and}
  \bibinfo{person}{Dani Shaket}.} \bibinfo{year}{2010}\natexlab{}.
\newblock \showarticletitle{Venus: Verification for untrusted cloud storage}.
  In \bibinfo{booktitle}{\emph{Proceedings of the 2010 ACM workshop on Cloud
  computing security workshop}}. \bibinfo{pages}{19--30}.
\newblock


\bibitem[\protect\citeauthoryear{Xu, Pokhrel, Lan, and Li}{Xu
  et~al\mbox{.}}{2022}]%
        {xu2022fair}
\bibfield{author}{\bibinfo{person}{Rongxin Xu}, \bibinfo{person}{Shiva~Raj
  Pokhrel}, \bibinfo{person}{Qiujun Lan}, {and} \bibinfo{person}{Gang Li}.}
  \bibinfo{year}{2022}\natexlab{}.
\newblock \showarticletitle{FAIR-BFL: Flexible and incentive redesign for
  blockchain-based federated learning}. In
  \bibinfo{booktitle}{\emph{Proceedings of the 51st International Conference on
  Parallel Processing}}. \bibinfo{pages}{1--11}.
\newblock


\end{thebibliography}


\end{document}